# Photoluminescence quantum efficiency of Er optical centers in GaN epilayers

V. X. Ho,[1] T. V. Dao,[1] H. X. Jiang,[2] J. Y. Lin,[2] J. M. Zavada,[3] S. A. McGill,[4] N. Q. Vinh[1,*]

[1]Department of Physics & Center for Soft Matter and Biological Physics, Virginia Tech, Blacksburg, Virginia 24061, USA
[2]Department of Electrical and Computer Engineering, Texas Tech University, Lubbock, Texas 79409, USA
[3]Department of Electrical and Computer Engineering, New York University, Brooklyn, New York 11201, USA
[4]National High Magnetic Field Laboratory, 1800 E. Paul Dirac Dr., Tallahassee, Florida 32310, USA

*Correspondence and requests for materials should be addressed to N.Q.V.; email:vinh@vt.edu

**We report the quantum efficiency of photoluminescence processes of Er optical centers as well as the thermal quenching mechanism in GaN epilayers prepared by metal-organic chemical vapor deposition. High resolution infrared spectroscopy and temperature dependence measurements of photoluminescence intensity from Er ions in GaN under resonant excitation excitations were performed. Data provide a picture of the thermal quenching processes and activation energy levels. By comparing the photoluminescence from Er ions in the epilayer with a reference sample of Er-doped $SiO_2$, we find that the fraction of Er ions that emits photon at 1.54 μm upon a resonant optical excitation is approximately 68%. This result presents a significant step in the realization of GaN:Er epilayers as an optical gain medium at 1.54 μm.**

The incorporation of rare earth elements into wide bandgap semiconductors is of significant interest for optoelectronic device applications, because of their temperature independent, atomic-like and stable emission together with the possibility of optical or electrical excitation[1, 2]. Much of the research has focused on the element Er with the emission from the first excited state ($^4I_{13/2}$) to the ground state ($^4I_{15/2}$) at 1.54 μm that is the minimum loss window of silica fibers for optical communications and in the eye-safe wavelength region[3-7]. GaN with the direct bandgap semiconductor appears to be excellent host materials for Er ions, not only due to their structural and thermal stability[8-10] but also to the recent advancements in growth techniques of high-quality crystals of both n- and p-type[2]. While light emitting diodes based on GaN:Er have been demonstrated[11], the realization of GaN:Er materials for optical amplification is still under investigation. For this reason, it is necessary to determine important factors which influence the optical performance of Er embedded in GaN. The information would provide us with the direction to optimize the optical properties of GaN:Er material.

Several important factors that determine the optical gain in GaN:Er need to unravel including the fraction of Er ions that contribute to light emission, number of optical centers as well as the energy transfer between the GaN host and Er ions. Previous work has revealed that a number of Er optical centers as well as a variety of energy transfer routes take place in GaN.[12, 13] The existence of various Er optical centers depends on preparation methods, such as ion implantation,[14, 15] metal-organic chemical vapor deposition (MOCVD)[16] and molecular beam epitaxy (MBE),[17] as well as growth and annealing conditions.[14] A single type of Er optical center was reported in GaN epilayers grown by MOCVD method.[18, 19] The absorption and emission studies together with the crystal field calculation provided an understanding of energy transfer mechanism to Er ions and the Er related luminescence process.[20] Non-radiative recombination channels were also investigated to understand photoluminescence (PL) quenching.[21] The Er-related trap centers with energy levels in the GaN bandgap act not only as active centers for bound excitons transferring their energy to 4f electrons of $Er^{3+}$ ions, but also as PL quenching centers.[10, 22, 23] These previous studies provided important insights for the improvement of Er emission in material engineering towards optimizing the energy transfer between the GaN host and Er ions. In our previous report, isolated Er and defect-related Er optical centers have been identified through high-resolution infrared PL spectroscopy.[24] Understanding the optical excitation mechanisms, optical activity of $Er^{3+}$ infrared luminescence and quenching channels is essential to enhance the emission efficiency of GaN:Er.

We present in this work a careful analysis of Er luminescence in GaN epilayers by comparing the PL intensity from our GaN:Er epilayers to a well-characterized standard reference sample of $SiO_2$:Er. The $SiO_2$:Er system has been used as an optical gain medium for solid-state lasers at 1.54 μm.[4, 25] The Er ions embedded in the defect free, insulating host matrix, $SiO_2$ reference sample, has an internal quantum efficiency (IQE) over 98%. Since basically all the Er ions in the $SiO_2$ matrix participate in photon emission, the reference sample can be employed as a benchmark to determine the optical activity of Er in other matrices. A low thermal quenching of 20% from 10 K to room temperature from the isolated Er optical centers in GaN epilayer at the 1.54 μm emission has been demonstrated. Employing a high resolution spectroscopy and temperature dependent measurements of PL intensity under resonant excitation provides an estimate of the quantum efficiency of PL process of Er optical centers as well as the thermal quenching mechanisms in GaN epilayers grown by MOCVD technique.

**Sample preparation and experimental methods**

The Er doped GaN epilayer samples were prepared by MOCVD method in a horizontal reactor.[10, 16, 26] A GaN:Er epilayer, with 0.5 μm thickness and Er concentration ($n_{Er}$) of ~ $10^{21}$ $cm^{-3}$ were grown on a thin un-doped GaN template of 1.2 μm on top of (1000) c-plane sapphire substrate. The growth temperature of Er-doped GaN layer was 1040 °C. The X-ray diffraction spectra indicated that GaN:Er epilayers have high crystallinity and no second phase formation. The band gap energy of GaN:Er epilayers is about 3.4 eV at room temperature. A detail description of the growth process and epilayer structure has been reported previously[10, 11, 16, 24, 26].

The high resolution PL spectra were conducted using a Horiba iHR550 spectrometer equipped with a 900 grooves/mm grating blazed at 1500 nm and detected by a high sensitivity liquid nitrogen InGaAs DSS-IGA detector. The resolution of PL spectrum is 0.05 nm. The PL experiments were carried out in a variable temperature closed-cycle optical cryostat (Janis) providing a temperature range from 10 K to 300 K. Both resonant excitation and the non-resonant excitation were employed to investigate the optical properties of Er in GaN epilayers.[24] The resonant excitation PL spectra from $^4I_{15/2} \rightarrow {}^4I_{9/2}$ of $Er^{3+}$ in GaN were obtained using a tunable wavelength Ti:Sapphire laser around 809 nm (1.533 eV) with a repetition rate of 80 MHz.[24]

The influence of $Er^{3+}$ site on the PL emission can be determined from optical excitation mechanisms. We have reported direct evidence of two mechanisms responsible for the excitation of optically active $Er^{3+}$ ions in GaN epilayers grown by MOCVD in our previous work.[24] Under resonance excitation via the higher-lying inner 4f shell transitions and non-resonant (band-to-band) excitation of the GaN host, the high resolution PL spectra at 10 K reveal an existence of two types of Er optical centers including the isolated and the defect-related Er optical centers in GaN epilayers.[24] For the first case, the isolated Er optical centers occupying Ga substitutional sites were observed under both the resonant ($^4I_{15/2} \rightarrow {}^4I_{9/2}$) excitation and the band-to-band excitation. Er ions in substitutional sites are considered as an isoelectronic impurity center.[27] The center, with no net charge in the local bonding region, can be excited by resonant and band-to-band excitation. Under the band-to-band excitation, a hole or an electron can be localized at the isolated center by a local core potential; subsequently, the secondary particle can be captured by Coulomb field of the first particle. The recombination of the two particles will transfer their energy to the Er ion. For the second case, the defect-related Er optical center can only be observed through the band-to-band excitation of the host involving a trapped (bound) exciton. The observation has been confirmed with a photoluminescence excitation measurement.[24] The excitation mechanism for the defect-related Er centers is believed to be related to intrinsic defects, impurities or defect-impurity complexes near the Er optical center. These defects capture excitons and subsequently transfer non-radiatively their energy to nearby Er ions. The efficiency of this process is high, but the requirement of bound excitons for excitation opens up non-radiative recombination channels for the luminescence process. At room temperature, we do not observe PL emission from the defect-related Er optical centers. For optoelectronic applications of the GaN:Er epilayers, this work focuses on the optical characterization of the isolated Er optical centers under the resonant excitation.

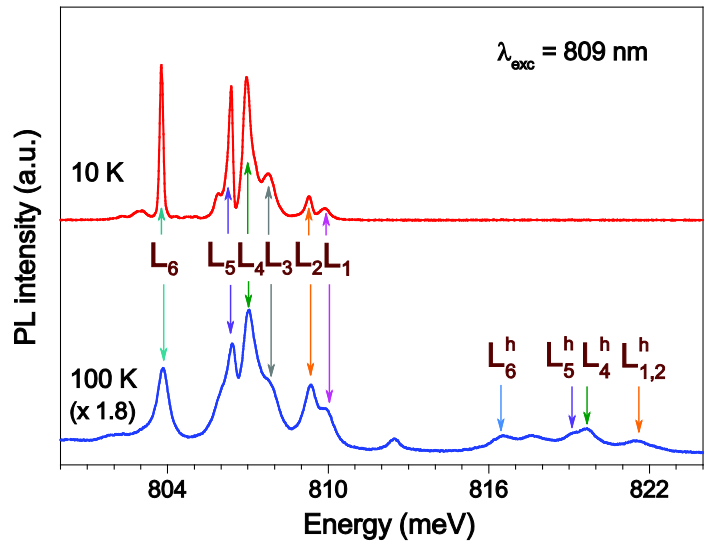

**Figure 1.** The PL spectra at 10 K and 100 K of the GaN:Er epilayer at 1.54 µm within the $^4I_{13/2} \rightarrow\ ^4I_{15/2}$ transition under the resonant excitation ($^4I_{15/2} \rightarrow\ ^4I_{9/2}$) excitation using a Ti:Sapphire laser ($\lambda_{exc}$ = 809 nm). At low temperature, the spectrum of the isolated Er optical center consists of a set of narrow and intense PL lines ($L_1$ to $L_7$). At higher temperatures, hotlines, $L^h_{1,2}$, $L^h_4$, $L^h_5$, $L^h_6$, appear and are displaced by about 12.6 meV. The intensities of hotlines rapidly increase with increasing temperature while the intensities of the main PL lines decrease.

**Results and discussion**

To determine the crystal field splitting of the first excited state ($^4I_{13/2}$) of the isolated Er optical centers, the high resolution spectra as well as the temperature dependence of PL spectra, have been investigated under resonant excitation (Fig. 1). At low temperatures, the spectrum of the isolated Er optical center consists of a set of narrow and intense PL lines (L center): $L_1$, $L_2$, $L_3$, $L_4$, $L_5$, $L_6$ and $L_7$ at energies of 809.88, 809.28, 807.80, 806.85, 806.32, 803.62, 802.91 meV, respectively. At higher temperatures, other PL lines, labeled hotlines $L^h_{1,2}$, $L^h_4$, $L^h_5$, $L^h_6$, appear at ~ 821.5, 819.6, 819.2 and 816.5 meV, respectively (Fig. 1). The PL intensity of PL lines $L_3$, $L_7$, are weak at low temperature and we cannot resolve the PL of hotlines $L^h_3$, $L^h_7$ at higher temperature. These hotlines are displaced by about 12.6 meV. At high temperature, the intensities of the hotlines rapidly increase with increasing temperature while the intensities of the main PL lines decrease. We note that due to the temperature broadening at high temperature the PL from $L_1$, $L_2$ lines merge into one PL line, thus we cannot resolve the PL hotlines $L_1^h$, $L_2^h$.

In order to determine the non-radiative transfer energy between $Er^{3+}$ ions and GaN host, we have measured the integrated PL intensity at the 1.54 µm emission as a function of temperature (Fig. 2). The integrated PL intensity measurements for the whole 1.54 µm band under the resonant excitation (λ = 809 nm) indicate a low thermal quenching of 20% from 10 K to room temperature from $Er^{3+}$ ions in our GaN:Er epilayer (Fig. 2, inset). These measurements determine the non-radiative recombination channels causing the thermal quenching of PL emission from isolated Er optical centers. In this model, the observed PL intensity for main PL lines of isolated centers should follow the Arrhenius equation:

$$I(T) = \frac{I_0}{1 + C_1 \exp\left(-\frac{E_{A1}}{k_B T}\right) + C_2 \exp\left(-\frac{E_{A2}}{k_B T}\right)} \tag{1}$$

where $I(T)$ is the integrated PL intensity at the temperature of T, $I_0$ is the integrated PL intensity at 10 K, $E_{A1}$ and $E_{A2}$ are the activation energies of the thermal quenching processes, $C_1$ and $C_2$ are fitting constants related to the density of non-radiative recombination centers in the sample, and $k_B$ is Boltzmann's constant. As shown in the Fig. 2, the solid line presents the best fit to the integrated PL intensity using Eq.1. The activation energy values derived from the fitting are $E_{A1}$ = 13 and $E_{A2}$ = 118 meV.

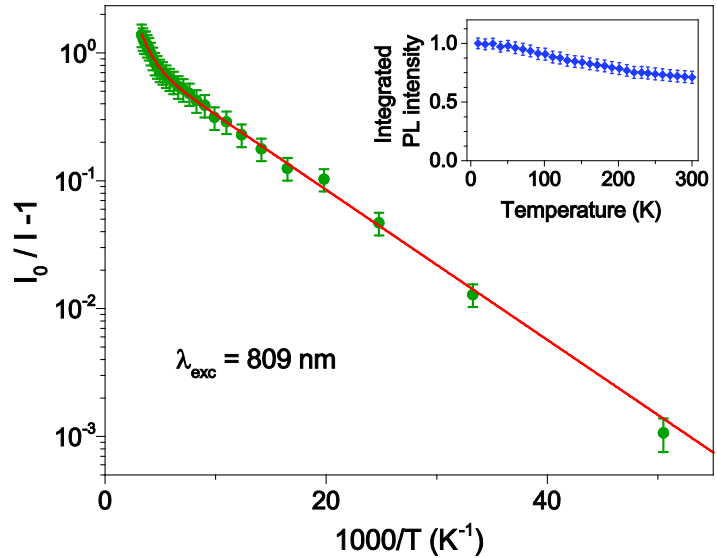

**Figure 2.** The temperature dependence of PL intensity measured under the resonant excitation ($\lambda_{exc}$ = 809 nm) indicates a low thermal quenching from $Er^{3+}$ ions in our GaN:Er epilayer. The plot illustrates the ($I_0/I$-1) vs. $T^{-1}$ behavior that was used to guide the Arrhenius fitting in which $I$ is the integrated PL intensity at the temperature of $T$, $I_0$ is the integrated PL intensity at 10 K. Activation energies of 13 and 118 meV have been determined. (**inset**) The integrated PL intensity measurements at 1.54 µm band show a thermal quenching of 20% from 10 K to room temperature from isolated Er optical centers.

Obviously, the replications of PL lines at higher temperature originate from the transition of the second-lying crystal field split level of the first ($^4I_{13/2}$) excited state to the sublevels of ($^4I_{15/2}$) ground state. When temperature increases, electrons from the lowest level of the first excited state gain energy and populate at the second-lying crystal field split level. Thus, the intensity of the main PL lines decreases and the intensity of hotlines increases with temperature. These hot lines are displaced by 12.6 meV, which is nearly equal to the activation energy of 13 meV. An energy level diagram for the luminescence of isolated Er optical centers is illustrated in the Fig. 3.

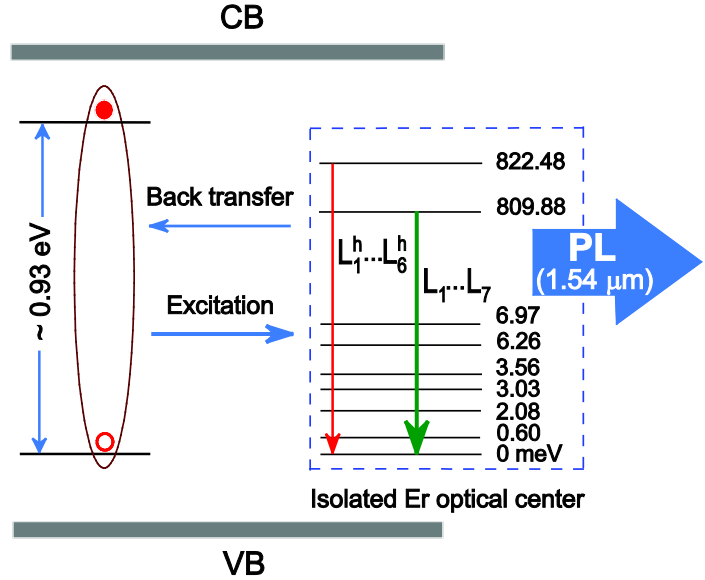

**Figure 3.** The Stark sub-energy level diagram of the excited and ground states of the isolated Er optical center and a deep defect level which are responsible for the luminescence thermal quenching process of $Er^{3+}$ ions.

A local density functional study[23] of Er doped GaN system showed that Er-related defect complexes with different local environments have different electronic properties for Er optical centers. For the 1.54 μm emission, deep defect levels associated with Er optical centers are expected to act as carrier traps. A hole or an electron can be localized at the deep level by a local core potential; subsequently, the secondary particle can be captured by Coulomb field of the first particle. The recombination of the two particles will transfer their energy to the Er ion. The reverse of this process, an energy "back transfer," is believed to be responsible for the decrease of PL intensity in these systems at elevated temperatures. Following the proposed excitation model, the energy back-transfer process in the isolated Er optical centers involves the de-excitation of excited $Er^{3+}$ by transferring their energy to the deep defect level of ~ 0.93 eV (Fig. 3). At the deep level, the energy deficit, in this particular case ~120 meV, is compensated by the annihilation of lattice phonons.[21, 28-30]. Such the deep defect level has recently predicted using the first-principles hybrid density functional study for Er-related defect complexes in GaN[31]. By optimizing the synthesis conditions including Er concentration, growth temperature, Er precursor vapor pressure, V/III flux ratio, we have reduced significantly the concentration of defect centers during the MOCVD growth and obtained strong PL intensity at room temperature.

The percentage of Er dopants that emit photons upon excitation is a crucially important parameter for the potential application of Er-doped semiconductors. In particular, it determines the PL emission efficiency at 1.54 μm of GaN:Er materials as well as the population inversion. An estimate of the number of emitting Er optical centers can be made by comparing the PL intensity of the GaN:Er epilayer grown by MOCVD technique with that of the $SiO_2$:Er reference sample with a similar shape and under the same experimental conditions. The experiment at room temperature has been performed under the resonant excitation $^4I_{15/2} \rightarrow {^4I_{9/2}}$ transition using the Ti:Sapphire laser at 809 nm. The time-integrated PL intensity of these samples has collected as a function of applied photon fluxes (Fig. 4). The instantaneous PL intensities of both samples were proportional to $N^*_{Er}/\tau_{rad}$, where $N^*_{Er}$ and $\tau_{rad}$ are the density of excited $Er^{3+}$ ions and their radiative lifetimes, respectively. Since the PL signal is integrated over time, the result of the experiment will be proportional to $N_{Er}^* \times \tau/\tau_{rad}$, where $\tau$ is the effective PL lifetime. The ratio of effective and radiative lifetime is the internal quantum efficiency (IQE), $\mathrm{IQE} = \tau/\tau_{rad}$, of the optical center. To obtain the number of photons coming out of the sample, the external quantum efficiency (EQE) is determined by the refractive index of the material for a specific wavelength, *i.e.*, $\mathrm{EQE} = \eta \times \mathrm{IQE}$, where $\eta$ is the extraction efficiency. In this calculation we ignore the photon reabsorption, therefore, the ratio of the number of photons emitted from the two investigated samples is given by:

$$\frac{I_{\mathrm{GaN:Er}}}{I_{\mathrm{SiO_2:Er}}} = \frac{\eta_{\mathrm{GaN}}}{\eta_{\mathrm{SiO_2}}} \frac{N^*_{\mathrm{Er(GaN)}}}{N^*_{\mathrm{Er(SiO_2)}}} \frac{\left(\tau_{\mathrm{eff}}^{\mathrm{Er(GaN)}}/\tau_{\mathrm{rad}}^{\mathrm{Er(GaN)}}\right)}{\left(\tau_{\mathrm{eff}}^{\mathrm{Er(SiO_2)}}/\tau_{\mathrm{rad}}^{\mathrm{Er(SiO_2)}}\right)} \quad (2)$$

The ratio of the extraction efficiencies can be calculated from the refractive indexes of GaN and SiO$_2$ materials.

$$\frac{\eta_{\mathrm{GaN}}}{\eta_{\mathrm{SiO_2}}} = \frac{n_{\mathrm{air}}^2/(4n_{\mathrm{GaN}}^2)}{n_{\mathrm{air}}^2/(4n_{\mathrm{SiO_2}}^2)} = 0.389 \quad (3)$$

where $n_{\mathrm{GaN}}$ and $n_{\mathrm{SiO2}}$ are the refractive indexes of 2.318 and 1.445 at 1.54 μm for GaN and SiO$_2$, respectively.

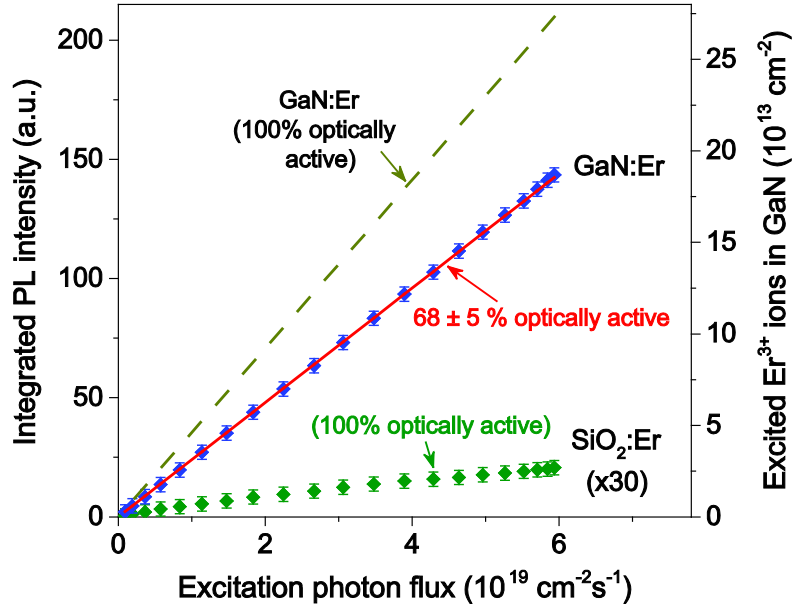

**Figure 4.** Comparison of the integrated PL intensity of Er$^{3+}$ ions in GaN epilayer with the reference SiO$_2$:Er sample under the resonant ($^4I_{15/2} \rightarrow {}^4I_{9/2}$) excitation indicates that a fraction of ~ 68% of Er$^{3+}$ ions in GaN sample are optically active.

Under the resonant excitation, the decay dynamics of all PL lines of the isolated Er optical centers appear as a single exponential with a decay time constant of 3.3 ± 0.3 ms.[24] The value is similar to those of previous reports on the Er in GaN.[12, 32, 33] The decay dynamics is effectively independent of temperature over the entire range from 10 K to room temperature. Thus, the radiative lifetime is 3.3 ms, and the internal quantum efficiency of the isolated Er optical center in GaN equals to 1, $\mathrm{IQE}_{\mathrm{Er(GaN)}} = \tau_{\mathrm{eff}}^{\mathrm{Er(GaN)}}/\tau_{\mathrm{rad}}^{\mathrm{Er(GaN)}} = 1$. For the SiO$_2$:Er standard sample, the Er embedded in the defect free, insulating host matrix, SiO$_2$, has an internal quantum efficiency (IQE) over 98%. Thus, we obtain $\mathrm{IQE}_{\mathrm{Er(GaN)}} = \tau_{\mathrm{eff}}^{\mathrm{Er(SiO_2)}}/\tau_{\mathrm{rad}}^{\mathrm{Er(SiO_2)}} = 1$.

The well-characterized SiO$_2$:Er sample can be used to attribute the measured PL intensity to a particular density of excited Er$^{3+}$ ions. The number of excited Er$^{3+}$ ions in SiO$_2$ reference sample can be calculated through excitation photon flux when we assume that all Er$^{3+}$ ions are equivalent and they all contribute to the PL process. As has been discussed previously,[6, 24, 34] under steady state conditions, the photon flux dependence of Er PL intensity is well described with the formula

$$I_{PL} \propto N^*_{Er(SiO_2)} = 100\% \times \frac{N^{ex}_{Er(SiO_2)} \sigma_{abs(SiO_2:Er)} \tau^{Er(SiO_2)}_{rad} \Phi}{1 + \sigma_{abs(SiO_2:Er)} \tau^{Er(SiO_2)}_{rad} \Phi} \quad (4)$$

where $\sigma_{abs(SiO_2:Er)} = 4.17 \times 10^{-22}$ cm$^2$ is the absorption cross-section of Er$^{3+}$ for the $^4I_{15/2} \rightarrow {^4I_{9/2}}$ transition,[4] $\tau^{Er(SiO_2)}_{rad} = 14.5$ ms is the radiative lifetime in the excited state, $^4I_{13/2}$,[4] $\Phi$ is the excitation photon flux, and $N^{ex}_{Er(SiO_2)} = 9.9 \times 10^{14}$ cm$^{-2}$ is the concentration of Er ions in the SiO$_2$ reference sample. As shown in Fig. 4, the photon flux dependence of PL intensity from SiO$_2$:Er sample shows a linear behavior for low excitation density. Under this condition, *i.e.* $\sigma_{abs}\tau_{rad}\Phi \ll 1$, this formula gives a linear dependence on the flux: $N^*_{Er(SiO_2)} = N^{ex}_{Er(SiO_2)} \sigma_{abs(SiO_2:Er)} \tau^{Er(SiO_2)}_{rad} \Phi$. Finally, since multiple reflections at the interfaces (air-GaN, GaN-air, air-SiO$_2$ and SiO$_2$-air) also might play a role, we have employed Fresnel's equations for calculations of the photon flux from the Ti:Sapphire laser entering the active layer. By substituting this into Eq. 2, we can estimate the number of excited Er$^{3+}$ ions in GaN sample (Fig. 4). We rescaled the right hand side scale of Fig. 4 until the solid red line, the calculated density of excited Er$^{3+}$ ions, overlaps with the PL intensity of GaN sample.

The PL from the GaN:Er sample also shows a linear dependence on the photon flux. As mentioned above, only the isolated Er optical centers are excited under resonant excitation at 809 nm. Thus, the excitation of the isolated Er optical centers in GaN can be calculated from the photon flux and the absorption coefficient under low excitation density:

$$N^*_{Er(GaN)} = A\% \times \left( N^{ex}_{Er(GaN)} \sigma_{abs(GaN:Er)} \tau^{Er(GaN)}_{rad} \Phi \right) \quad (5)$$

where A is the percentage of Er ions that are isolated Er optical centers in GaN epilayer, $\sigma_{abs(GaN:Er)} = 3 \times 10^{-20}$ cm$^2$, $\tau^{Er(GaN)}_{rad} = 3.3$ ms,[24] $N^{ex}_{Er(GaN)} = 1 \times 10^{21}$ cm$^{-3}$, and the thickness of GaN:Er epilayer is 500 nm.[35] The dash line represents the case that if all Er ions in the epilayer were optically active, A = 100%. From this calculation we determine that the percentage of the isolated Er optical centers in epilayer, which is ~ 68 ± 5 %. The uncertainty of the value is determined by the fluctuation in the photon flux and the measurement temperatures. The percentage of isolated Er optical centers is high enough that we can expect optical amplification in these materials. This value is higher than that of Eu optically active centers in GaN materials emitting at about 620 nm that had been determined by a similar approach.[7]

**Conclusions**

In summary, we have investigated the optical activity, the PL quantum efficiency and non-radiative transfer energy pathways between Er ions and GaN host from GaN:Er epilayers prepared by MOCVD. The PL intensity measurements under the resonant excitation ($\lambda$ = 809 nm) indicate a low thermal quenching of 20% from 10 K to room temperature from Er$^{3+}$ ions in our GaN:Er epilayer. By comparing the PL intensity from the GaN epilayer with that of the SiO$_2$:Er reference sample under the same conditions, we estimate the percentage of isolated Er optical centers in the GaN epilayer is approximately of 68%. The high percentage of optically active centers in GaN epilayers indicates the high potential for realizing optical amplification in GaN:Er materials grown by MOCVD. Employing the temperature dependence measurements of the PL intensity, we also have identified non-radiative channels in this material. The findings provide useful insights for further improvement of the 1.54 μm emission in material engineering towards optimizing the energy transfer between GaN host and Er ions.

**Acknowledgments**

N.Q.V. acknowledges the support from NSF (ECCS-1358564). The materials growth effort at TTU was supported by JTO/ARO (W911NF-12-1-0330). The authors gratefully acknowledge F. W. Widdershoven for preparation of the $SiO_2$:Er reference sample used for calibration.